\documentstyle[preprint,pre,aps,eqsecnum]{revtex} 
\input{epsf} 
\begin{document} 
\bibliographystyle{normal} 
%\tightenlines  

\draft 

\title{Distribution of injected power fluctuations in electroconvection}  
\author{Tibor T\'oth-Katona{\footnote{On leave from: 
Research Institute for Solid State  
Physics and Optics, Hungarian Academy of Sciences, 
H-1525 Budapest, P.O.B.49, Hungary }} and J. T. Gleeson}  

\address{Department of Physics, Kent State University,  
P.O.B. 5190, Kent, OH 44242, USA } 
\date{\today} 
\maketitle 

\begin{abstract} 
We report on the distribution spectra of the fluctations in the 
amount of power injected into a liquid crystal undergoing electroconvective 
flow.  The probability distribution functions (PDFs)  of the 
fluctuations as well 
as the magnitude of the fluctuations  have been determined in a wide range of 
imposed stress both  for `unconfined' and `confined' flow geometries. 
These spectra are compared to those found in other systems held 
far from equilibrium, and find that in certain conditions we obtain 
the ''universal'' PDF form reported in [Phys. Rev. Lett. {\bf 84}, 3744 (2000)]. 
Moreover, the PDF approaches this universal form via an interesting mechanism
whereby the distribution's negative tail evolves towards form in
a different manner than the positive tail.

\end{abstract}  


Fluctuations in systems driven out of equilibrium have recently attracted  
considerable attention, particularly with regard to the probability density 
function (PDF) of fluctuations in global quantities.  
Fluctuations in global quantities are necessarily  
the result of many individual fluctuating modes,  thus the first issue 
is whether the central limit theorem,  
which predicts a Gaussian PDF, holds. Recent results in a 
number of disparate systems  reveal non-Gaussian 
PDF's exhibiting rich and intriguing behavior. 
Furthermore, the understanding of 
such PDF's is of 
practical importance,  not least because one would like to predict the probability 
of exceedingly rare fluctuations having  colossal  
amplitude (e.g. floods, violent storms, earthquakes, stockmarket swings).  
While non-Gaussian PDF's of fluctuations are intriguing in  
their 
own right,  recent results suggest there may exist a universal, 
non-Gaussian distribution of global fluctuations.  Strikingly,  such 
a distribution has been  
found, using no adjusted parameters or fits, for an astonishing variety of seemingly 
unrelated systems: turbulent flow in confined 
geometry,\cite{cadot95,labbe96,labbe96a,mordan97,bramw98,pinton99,auma00,auma03},  
the  Danube water level,\cite{bramw02}  and 
simulations of the 3D X-Y model at criticality \cite{bramw98,bramw00,auma01}.  In 
all these systems, the PDF is substantially skewed, with one tail well described by an 
exponential decay. 
This distribution  is well described by generalized 
Fisher-Tippet-Gumbel (gFTG)  distribution \cite{bramw00}.  The exponential tail is 
adduced\cite{bramw98,pinton99,porte03}  to be due to fluctuations  
having length scale comparable to the system size. This 
explanation is supported by measurements on turbulent swirling flow in 
unconfined geometry \cite{labbe96}, in which no 
exponential tail has been found and the  fluctuations became Gaussian.   
Note that all the above listed results have been obtained 
for  {\it isotropic} fluids.    

For flow of 
{\it anisotropic} fluids, velocity fluctuations of tracer 
particles  have been investigated\cite{kai02} in 
the so-called soft mode turbulence  \cite{richt95}, and with the increase of 
the stress a change in PDF  has been observed from L\'evy to Gaussian 
via some intermediate  distributions such as the exponential one.  However, it should be 
borne in mind that these represent local rather than global 
measurements.  Electrohydrodynamic convection (EHC) in liquid crystals (LCs)  is a 
unique system in which abrupt turbulence to turbulence  transitions [such as 
defect turbulence to dynamic scattering mode 1  (DSM1) or DSM1 
to DSM2] occur at well defined thresholds.  The study of 
the average injected power and fluctuations in  quantity in EHC has been 
established in Refs. \cite{kai77,jim01,gold01}.   
This method opens new routes in investigations of EHC. In this 
Letter we analyze the PDF of fluctuations in an anisotropic fluid system 
driven far out of equilibrium. 
EHC affords the opportunity of 
varying the externally imposed stress over a sufficiently wide range that it 
is possible to observe the evolution of the PDF shape.  Furthermore, 
our system allows detailed studies of the effects of confinement on the PDF evolution. 
The latter is important because the experimental results of Ref. \cite{labbe96} show  
substantial, qualitative differences between PDF forms for fluctuations of global injected 
power  in `unconfined' and in `confined'  geometries.    

In turbulent swirling flow experiments in which the fluctuations in injected power 
are measured, the stress applied to the fluid is characterized by 
the Reynolds number ($Re$). The comparison of PDF between 
confined and unconfined flow was made over a range of $Re$ less than 10 \cite{labbe96}. 
Direct comparison with  EHC is problematic because the stress applied to the 
LC inducing flow is characterized by 
not by a Reynolds number but rather by the dimensionless potential 
difference, $\varepsilon \equiv U^2/U_c^2 - 1$, 
where  $U$ is the applied potential difference  and $U_c$ 
is the critical potential difference necessary to induce flow.  
Two important advantages of EHC are,  
the ability to widely vary  both the relevant length scales and $\varepsilon$.   

Our experimental  setup is described in Ref. \cite{jim01}. A 
sinusoidal voltage signal is amplified and applied  across the LC layer sandwiched 
between two glass plates.  The current traversing through the LC sample 
returns to ground  via the field-effect transistor input of a 
current-to-voltage preamplifier.  The output of this preamplifier 
is measured by a lock-in amplifier whose reference signal is 
is supplied by the original function  generator. The in-phase output of the  
lock-in is amplified and digitized. For each experimental point an 
optical image taken through a polarizing  microscope with shadowgraph technique has 
been also recorded.  The liquid crystalline mixture Mischung V (MV) 
with $2.73wt \%$ dopant has been used which is 
an excellent model  material because of its chemical stability and known material 
parameters  \cite{shi02}. All the measurements presented below have been carried out 
at  temperature $T=(50.00 \pm 0.01)^{\circ}C$,  where a satisfactory spatial 
homogeneity of the sample is  ensured \cite{toth03}.   The LC is encased in sandwich-type 
cells with planar orientation  for both `unconfined', and `confined' flow geometry.  
For `unconfined' flow geometry, we chose a  cell with square, etched  electrodes having active 
area $A=(6.15 \pm 0.1) mm^2$ and thickness of  $d=(33.4 \pm 0.2)\mu m$. 
In this geometry, the electric field is present and the convection takes place within 
the active area. This area is laterally bounded by the remainder of the LC, thus 
the flow and director fields are not controlled at these boundaries. For the `confined' 
flow geometry,  a mylar gasket with a circular hole [$A=(24.6 \pm 0.6) mm^2$] was 
used to confine the LC between the conductive plates and within the active area.  
The separation between the plates was $d=(80 \pm 20)\mu m$.   The above described 
dimensions provide aspect ratios  $s = \sqrt{A}/d \approx 74$ 
for the `unconfined' flow geometry and  $s \approx 62$ for 
the `confined' cell. These values of $s$ are similar enough  to 
make quantitative comparison for injected power fluctuations  between the 
`unconfined' and `confined' geometry, knowing that the normalized  
variance of power fluctuations depends strongly on $s$ \cite{toth03a}. 
Before performing fluctuation measurements, the experimental setup 
was  tested by replacing the LC sample with 100M$\Omega$ ohmic 
resistor  (resistivity of the same order of magnitude 
as our samples).  Fluctuations in the current injected 
into the test resistor obey 
Gaussian statistics with $\sigma_P/ \langle P \rangle ~ < ~ 10^{-5}$. 

Figure 1 shows temporal dependence of the normalized power fluctuations  
around the mean value $\langle P \rangle$ for 
both unconfined and confined LC electroconvective flow at  
moderate stress: $\varepsilon = 42$.  
We want to emphasize two features of these fluctuations.  First, the normalized 
variance of fluctuations  
$\sigma_P /\langle P \rangle =\sqrt{(P-\langle P \rangle)^2}/\langle P \rangle$  
is of the same order of magnitude for unconfined and confined flow. 
We do not witness the significant increase in  
$\sigma_P/ \langle P \rangle$ when the flow is
confined as described in 
Ref.  \cite{labbe96}.    
Second, there is a qualitative difference between 
the power  fluctuations in the two flow geometries. 
For unconfined flow (at this value of $\varepsilon$) injected power 
fluctuations  are uniform, resulting in almost Gaussian PDF (see below). 
In contrast, during confined flow,  we observe relatively rare but 
intermittent fluctuations having large, negative amplitude 
(at least 6.5 standard deviations); 
these negatively skew the PDF which appears to be well described by the 
gFTG.   

Before discussing the forms of PDF, it is useful to summarize the results  
of the optical observations (performed concomitantly with the injected 
power  fluctuation measurements). In general, the EHC patterns have  similar 
appearance in unconfined and in confined flow geometry however, they appear 
at somewhat  different values of $\varepsilon$ for the two 
geometries and  they differ in details within the defect 
turbulence regime  (e.g., the grid pattern is observed in unconfined flow 
but not in confined flow). 
In both geometries, as $\varepsilon$ is increased above zero the stationary, oblique roll pattern 
appears. Defect turbulence (described more in detail below)  starts at 
$\varepsilon \approx 0.2$ for both flow geometries.  Defect turbulence 
is characterized by low-frequency, persistent  oscillations in the 
autocorrelation function $g_a(t)$ of the power  
fluctuations \cite{toth03a}.  The transition threshold from defect 
turbulence to DSM1 is defined as the  voltage $\varepsilon$ at 
which the persistent  oscillations in $g_a(t)$ diminish \cite{toth03a}. 
This transition occurs  at $\varepsilon \approx 7.7$ and at 
$\varepsilon \approx 12.5$  for the unconfined and confined flow, 
respectively. The DSM1 $\rightarrow$ DSM2  turbulence transition 
(involving an abrupt increase in density  of 
disclination loops) has been detected at  
$\varepsilon \equiv \varepsilon_t \approx 62$ and 
at $\varepsilon_t \approx 19.9$  for the unconfined and confined 
flow, respectively.  With 
further increase of $\varepsilon$ no more transitions 
are reported in the  literature.  This is unsurprising
because above $\varepsilon \approx 800$ 
the flow becomes so turbid that is
is impossible to visually detect any further change in the pattern.    

Figures 2 and 3 show PDFs of 
injected power fluctuations $\pi(P)$ scaled with  their variance 
$\sigma_P$ as a function of power around its mean 
value  $\langle P \rangle$ normalized with $\sigma_P$ at 
different imposed stresses  covering a range of about $10^3$ 
for both unconfined (open symbols) and  confined flow (closed symbols).  
The full lines are Gaussian  distributions as denoted in Fig. 2(a) 
with the same $\sigma_P$ as experimental results  (not fits). 
The dashed lines are the gFTG distribution:
\begin{equation}
%\pi (P) \sigma_P = K[e^{b(x-c)-e^{b(x-c)}}]^{a}
\pi (P) \sigma_P = K \exp(b(x-c)-e^{b(x-c)})^{a}
\end{equation}
where 
$x=(P- \langle P \rangle)/\sigma_P$, $K=2.14$, $a=\pi/2$, $b=0.938$ and 
$c=0.374$.
This is {\em not} a fit: all parameter values are taken from 
\cite{bramw00}.  This is the distribution referred to as ``universal'' 
in Ref. \cite{bramw00}.  

Slightly above EHC threshold, 
at $\varepsilon \approx 0.2$ the process  of generation/annihilation 
of defects (dislocations) starts which destroys  the 
stationary EHC roll pattern by breaking the rolls into moving segments  
and leading to a state called defect turbulence \cite{coul89}. 
Defect  turbulence causes dramatic increase in the amplitude of power 
fluctuations  and the fluctuations become quasi-periodic with a dominant 
frequency corresponding to the defect lifetime \cite{toth03a}.  
These fluctuations are well described by Gaussian distribution for both the 
unconfined and confined flow geometry; see Fig. 2(a).   

With further increase of $\varepsilon$, the PDF for 
{\it unconfined}  flow remains Gaussian  even above the 
defect turbulence $\rightarrow$ DSM1 transition -- see  
Figs. 2(b) and 2(c).  
Figure 2(d) shows PDFs obtained at   
$\varepsilon = 42$ (corresponding to Fig. 1).  In the  unconfined flow, 
we are deeply  in DSM1, and at this $\varepsilon$ the first systematic  departure 
from the Gaussian distribution is observed with tails on  decaying slower than 
Gaussian on both sides of the PDF.  With further increase of $\varepsilon$, 
but still staying in DSM1 turbulence  the deviation from the 
normal distribution becomes even more pronounced  [Fig. 3(a)].   At 
and above $\varepsilon_t \approx 62$, the PDF for unconfined flow 
abruptly reverts to Gaussian [Fig. 3(b)] and remains so for 
$\varepsilon$  up to about 860.  
Above this value, the PDF deviates again from Gaussian 
and its form is much closer to 
gFTG.\cite{bramw00} (c.f.  Fig. 3(c)); the PDF keeps this shape 
for extremely high $\varepsilon >1000$, [Fig. 3(d), open  
symbols, $\varepsilon=1717$].   

In stark contrast, in the {\it confined} flow geometry,  a systematic  
deviation from the Gaussian distribution is detected even in the 
defect  turbulence regime, above $\varepsilon \approx 4$ [closed 
symbols in Fig. 2(b)].  This deviation reminds us 
of the results obtained for swirling flow in confined 
geometry\cite{labbe96}.  Clearly, the negative tail of PDF for confined 
flow in Fig. 2(b) is  exponential and is in agreement with gFTG 
distribution.  The positive tail however, remains Gaussian.  Thus,  
at this range of stress we observe a ``hybrid'' distribution having gFTG 
tail for negative fluctuations but a Gaussian tail for positive fluctuations. 
The deviation from Gaussian distribution (and convergence to gFTG)  
is even more expressed above the defect  turbulence $\rightarrow$ 
DSM1 transition [closed 
symbols in Fig. 2(c)], where the positive tail also starts to  
approach the gFTG distribution.  In contrast to the unconfined flow geometry, 
in confined flow DSM1 $\rightarrow$ DSM2  transition has 
no noticeable influence on the form of 
PDF [{\it cf.}  Figs. 2(c) and 2(d)] which stays close to 
gFTG distribution up to  $\varepsilon \approx 1000$ 
(over a range of $O(10^3)$ of imposed stress)  
[Figs. 2(d) and 3(a)--(c)].  At 
extremely high stresses ($\varepsilon >1000$) however,  the form of PDF 
changes and the typical shape is shown in Fig. 3(d) 
(closed symbols, $\varepsilon=1424$)  with heavy tails on both 
negative and positive sides.      

The Gaussian PDFs in Fig 
2(a) for both confined and unconfined  flow suggest that 
fluctuations in global injected power  arise from many spatially 
uncorrelated contributions (defects). However, despite the spatial 
uncorrelation in defect turbulence,  there is still a 
surprising degree of temporal order 
embedded\cite{toth03,toth03a}.   
Figs. 2(b) and 2(c) are in the agreement with the results of  
Ref. \cite{labbe96} for swirling flow: for unconfined flows  
the PDF is Gaussian, for confined flow however, it is much 
closer to gFTG  distribution. The PDF for confined flow 
stays remarkably close to the gFTG distribution 
for $\varepsilon$ varying over $\approx 10^3$ 
[Figs. 2(b)-(d) and 3(a)-(c),  closed symbols]. Above 
$\varepsilon \approx 1000$ however,  the PDF changes; rare events 
of large amplitude fluctuations no longer follow the gFTG distribution; 
instead they form the above mentioned PDF with  
heavy tails [Fig. 3(d)]. For unconfined flow, the PDF remains 
Gaussian over a  range of $\varepsilon > 10^2$ except in a narrow 
range of $\varepsilon$ just below the  threshold of 
DSM1 $\rightarrow$ DSM2 turbulence transition 
[Figs. 2(d) and  3(a)]. At high $\varepsilon$ however, PDF of 
unconfined flow also follows gFTG  distribution [Figs. 3(c)-(d)].  
Furthermore, the skewness of the measured distributions is  $-(0.91 \pm 0.2)$ 
and $-(1.0 \pm 0.1)$ for  unconfined ($\varepsilon > 860$) 
and confined ($5 < \varepsilon < 1000$), which compares variably with 
the expected value  for the gFTG of -0.893.     

The results discussed above have several implications. First, the 
gFTG distribution  of global fluctuations is observable even in unconfined 
flow, but at substantially larger imposed stress. This suggests even 
more strongly that this distribution may be a universal trend for 
strongly fluctuating non-equilibrium systems, 
whether the 
flow is confined or not.
Second, when the imposed stress is sufficiently increased,  we 
observe departures from the gFTG distribution.  Thus,  this distribution, 
while exhibiting indications of being universal 
(in that the same form is observed for disparate systems) cannot be 
thought of as a limiting, ultimate shape. 
Interestingly, we have shown that the form of PDF is not simply dependent 
on the boundary conditions and  the applied stress, but 
in our system also depends on turbulence to  
turbulence transition(s) [c.f. Figs. 3(a) and (b)].   Of 
course,  such transitions are not common. 
Last, the confined flow experiments above reveal a fascinating 
mechanism whereby the 
PDF transforms as the stress is increased. 
Starting with Gaussian at low stress, the PDF morphs into a hybrid distribution in which the 
negative fluctuations follow the exponential decay of the gFTG, while the 
positive fluctuations remain Gaussian.  As the stress increases, the positive 
fluctuations then decay more quickly than Gaussian,  and the gFTG is obtained. 
We are unaware of any theoretical explanations of 
such transitions between PDF's.  

\begin{acknowledgements}  Financial support from the  National Science 
Foundation, Grant DMR-9988614 is kindly acknowledged.  We 
have benefited from discussions with W. Goldburg and Z. R\'acz. 

\end{acknowledgements}

\begin{figure}[h] 
\begin{center} 
\parbox{16.5cm}{ 
\epsfxsize=16cm 
\epsfbox{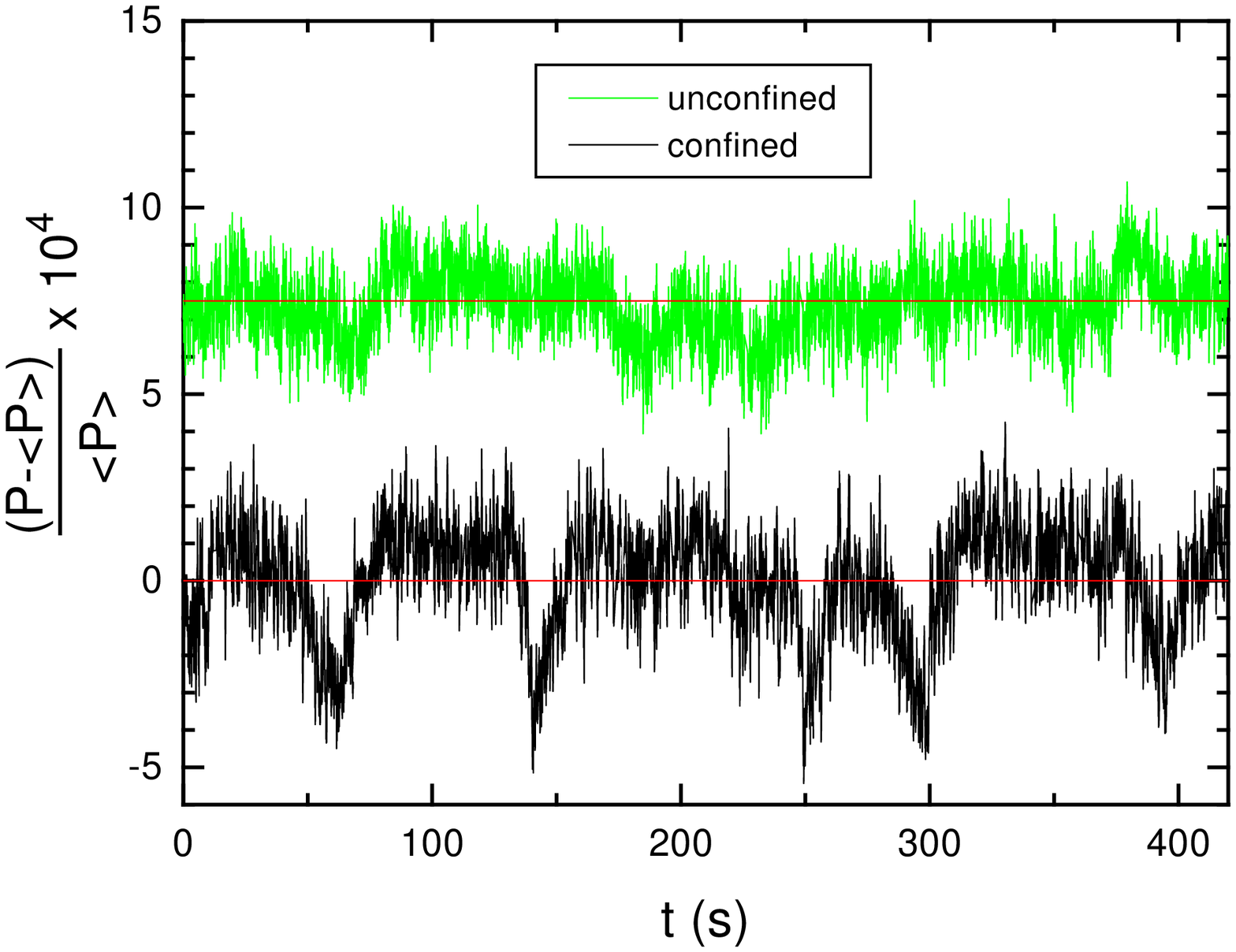}} 
\end{center} 
\caption{Temporal dependence of normalized injected power fluctuations  
around the mean value $\langle P \rangle$  
for unconfined flow (offset by $7.5 \times 10^{-4}$) 
and for confined flow geometry  at $\varepsilon=42$.  
} 
\label{tothfig1} 
\end{figure}  

\begin{figure}[h] 
\begin{center} 
\parbox{16.5cm}
{ 
\epsfxsize=16cm 
\epsfbox{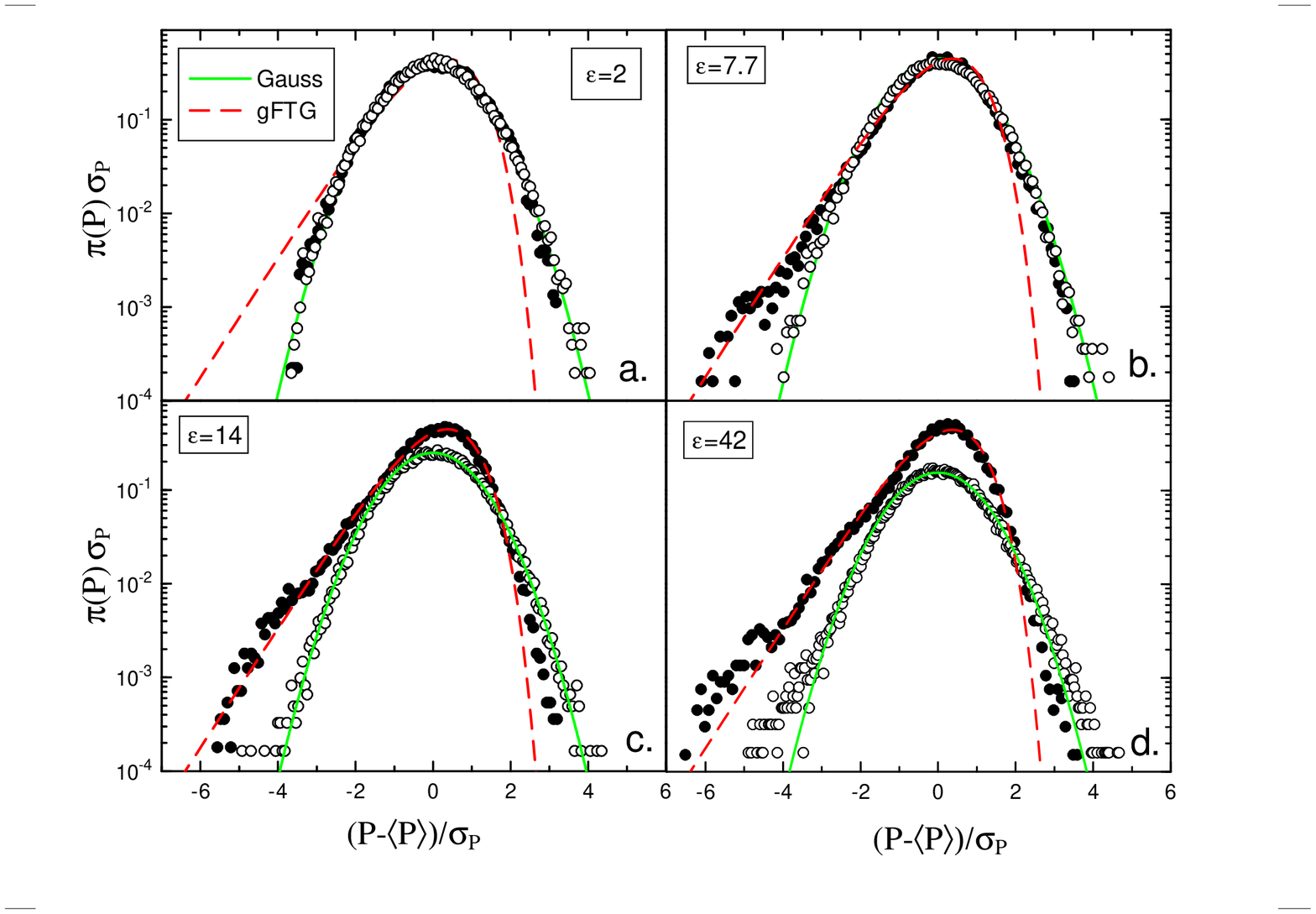}} 
\end{center} 
\caption{Probability density function for both unconfined 
(open symbols)  and confined flow geometry (closed symbols). 
Lines show Gaussian and gFTG distributions as denoted in legend.
(a) $\varepsilon=2$ -- defect turbulence for both geometry;   
(b) $\varepsilon=7.7$ -- DSM1 for unconfined flow and  defect turbulence for 
confined flow;  
(c) $\varepsilon=14$ -- DSM1 for both unconfined and confined flow;  
(d) $\varepsilon=42$ -- DSM1 for unconfined flow and DSM2 
for confined flow.  
} 
\label{tothfig2} 
\end{figure}  

\begin{figure}[h] 
\begin{center} 
\parbox{16.5cm}{ 
\epsfxsize=16cm 
\epsfbox{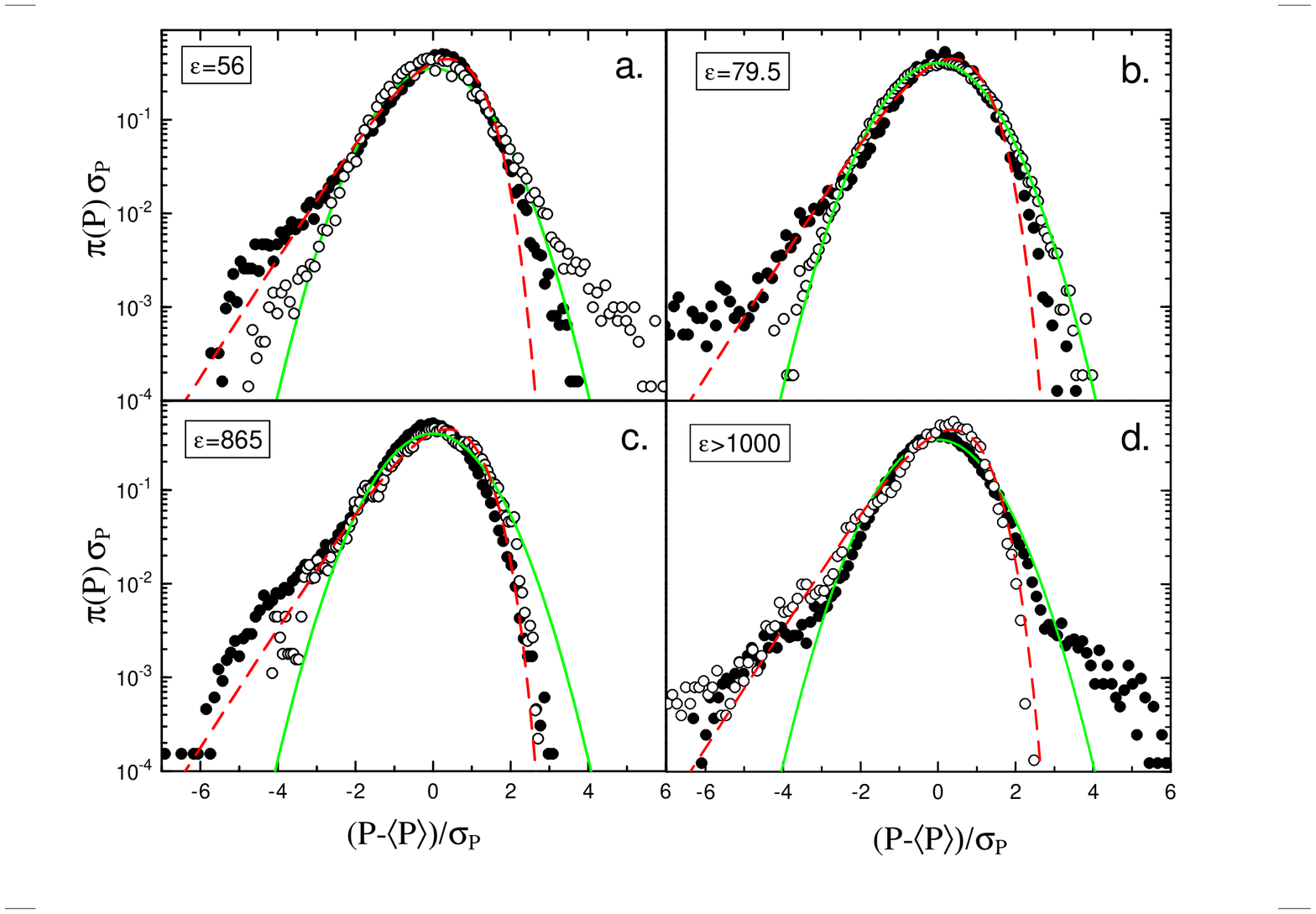}} 
\end{center} 
\caption{Same as Fig. 2, at higher values of $\varepsilon$.  
(a) $\varepsilon=56$ -- DSM1 for unconfined flow and DSM2 for confined flow;  
(b) $\varepsilon=79.5$ -- DSM2 for both unconfined and confined flow;    
(c) $\varepsilon=865$ -- DSM2 for both unconfined and confined flow;  
(d) $\varepsilon > 1000$ -- DSM2 for both unconfined and confined flow. 
} 
\label{tothfig3} 
\end{figure}

\end{document}